\def\Re{{I\!\! R}}
\def\fs{\footnotesize}

\def\al{\alpha}
\def\bt{\beta}

\def\Ga{\Gamma}
\def\de{\delta}

\def\ve{\varepsilon}

\def\la{\lambda}

\def\vr{\varrho}
\def\si{\sigma}

\def\om{\omega}
\def\Om{\Omega}
\def\th{\theta}

\def\vp{\varphi}

\def\pt{\partial}
\def\na{\nabla}
\def\w{\wedge}
\def\A{\hbox{\bf A}}
\def\F{\hbox{\bf F}}
\def\bg{\hbox{\bf g}}
\def\G{{\bf \Gamma}}
\def\R{\hbox{\bf R}}
\def\J{\hbox{\bf J}}
\def\tr{\hbox{\tt tr}}
\def\bx{\bar x}
\def\bchi{\bar \chi}

\documentstyle[11pt]{article}
\begin{document}

\title{LAGRANGIAN SYMMETRIES \\
OF CHERN--SIMONS THEORIES}

\author{
Andrzej BOROWIEC\thanks{On leave  from the
Institute of Theoretical Physics, University of Wroc{\l}aw,
pl. Maksa Borna 9, PL-50-204  WROC{\L}AW, POLAND
(e-mail: borow@ift.uni.wroc.pl).
}\\
 Marco FERRARIS\\
Mauro FRANCAVIGLIA\\
{\fs Dipartimento di Matematica}\\
{\fs Universit\`a di Torino}\\ {\fs Via C. Alberto 10, 10123 TORINO (ITALY)}
}
\date{July 15, 1998}

\maketitle

\begin{abstract}
This paper analyses the Noether symmetries and the corresponding conservation
laws for Chern-Simons Lagrangians in  dimension $d=3$. In particular, we
find an expression for the superpotential of Chern-Simons gravity. As
a by-product  the general discussion of superpotentials for 3rd order
natural and quasi-natural theories is also given.\\\bigskip
PACS number(s): 04.20.Fy, 11.10Kk, 11.30.-j, 04.20.Cv
\end{abstract}

\section{Introduction}

The interplay between symmetries, conservation laws and variational principles
for Lagrangian dynamical systems have been under consideration for a long
time, starting from the celebrated work by E. Noether. Our aim in this paper
is to calculate conserved quantities for Chern-Simons gravity in
dimension $d=3$. Chern-Simons theories exhibit in fact many interesting and
important properties. First of all, they are based on secondary characteristic
classes discovered in \cite{CS} and there is some hope that they will give  new
topological invariants for knots and three-manifolds. Moreover, they appeared
in physics as natural mass terms for gauge theories and for gravity in
dimension three, and after quantization they lead to a quantized coupling
constant as well as a mass \cite{DJT}.
They have also found  applications to a large variety of physical 
problems \cite{des}, among which we quote anyons and quantum Hall 
effect; when the gauge group is the inhomogeneous Lorentz group then such 
a theory is equivalent to standard gravity.
Chern-Simons gauge theory is also an example of a topological field theory
\cite{Wi}. It should also be remarked that, as it
has been recently shown, Chern-Simons Lagrangians are the only
obstructions to an equivariant inverse problem of the calculus of variations 
in $d=3$ \cite{An}. Finally, the Chern-Simons term is related to the
anomaly cancellation problem in $d=2$ conformal field theories \cite{CF}.

Our paper is organized as follows. For our later
convenience, we shall start from a short, elementary and self-contained 
exposition of a variational approach to the problem of conservation laws 
in field theories. A more systematic approach can be found e.g. in 
\cite{Go,Fl,BCJ,heh,JS}, while for a rigorous mathematical presentation 
\footnote{It involves a variational calculus on jet bundles.}
we refer the reader to \cite{FF1,FF2,FFF,sar}. In section 3 we collect 
some material from our earlier papers concerning superpotentials and 
in particular we shall list the explicit formulae which are necessary 
to calculate them in all theories of order at most three. Section 4 
deals with Chern-Simons gauge Lagrangians and we discuss how general 
formulae for superpotentials apply in this case.  Some results of 
section 4 are finally used in section 5 to calculate the 
energy-momentum complex for Chern-Simons gravity.

\section{Preliminaries}

Let us consider a field theory over a spacetime manifold $M$ of 
arbitrary dimension $d$, describing the behaviour of an arbitrary 
field $\phi$ by means of a Lagrangian  $L\equiv L(\phi, \pt_x\phi,\ldots)$ 
of arbitrary order. The action functional will be
$$
I(\phi)=\int_D L dx 
$$
where the integration  is performed over a domain $D$ of the
manifold $M$ of independent variables $x\equiv (x^\mu)$; the field
configuration (i.e., the set of all dependent variables) is represented 
here by $\phi$. \footnote{
For simplicity we drop an internal field index, e.g. $\phi^A$}
Variation (i.e. a functional derivative) $\de I$ of $I$ with respect 
to an arbitrary (infinitesimal) variation $\de\phi$ of $\phi$ leads 
to the variation $\de L$ of the Lagrangian itself. This, in turn, can 
be rearranged into two parts according to the "first variation formula"
\footnote{Here the usual Einstein summation convention over repeated
 indices is adopted.}
$$
\de L= \frac{\de L}{\de\phi} \de\phi + \pt_\mu \vr^\mu \eqno (1)
$$
The first term consists of the Euler-Lagrange expression
(field equations):
$$
\frac{\de L}{\de\phi}=\frac{\pt L}{\pt\phi} - \pt_\mu (\frac{\pt L}
{\pt\phi_\mu})+\pt_\mu\pt\nu (\frac{\pt L}{\pt\phi_{\mu\nu}})-\ldots
$$ 
and therefore vanishes on shell,
i.e. when the field $\phi$ satisfies the equations of motion.
The second part (boundary term) is a divergence of 
$\vr^\mu\equiv\vr^\mu (\phi, \de\phi)$, where
$$
\vr^\mu=[\frac{\pt L}{\pt\phi_\mu}-\pt_\nu (\frac{\pt L}
{\pt\phi_{\mu\nu}})]\de\phi + \frac{\pt L}{\pt\phi_{\mu\nu}}
\de\phi_\nu + \ldots \eqno (2)$$
This boundary term is usally neglected since it does not contribute 
to the equations of motion (and it is killed on the boundary $\pt D$ after 
imposing  suitable boundary conditions on $\de\phi$).  However, this second 
contribution is physically important: in fact, it does contribute to 
conservation laws.

Recall that a variation, say $\de_*\phi$, is called an {\it (infinitesimal)
symmetry} \ \  of the action $I$ if the corresponding variation $\de_* L$ can
be written as a divergence (i.e., $\de_* L=\pt_\mu\tau^\mu$) without using the
equations of motion. Of course, this  depends heavily on  the transformation
properties of $L$: for example, if the Lagrangian transforms as a scalar
(i.e. $L(\phi)=L(\phi +\de_*\phi)$), then one has $\de_*L=0$.

Therefore, for the variation implemented by a symmetry transformation
$\de_*\phi$, equation (1) can be rewritten under the following form
 $$
 \frac{\de L}{\de\phi} \de_*\phi = - \pt_\mu (\vr_*^\mu -\tau^\mu) \eqno (3)
$$
where $\vr_*^\mu = \vr^\mu (\phi, \de_*\phi)$. A {\it Noether current}
then arises
$$E^\mu\equiv E^\mu (\phi, \de_*\phi) = \vr_*^\mu -\tau^\mu  \eqno (4)$$
which is conserved on shell. One writes
$$
\pt_\mu E^\mu=0 \ \ \ \hbox{mod}\ \ \ \ \frac{\de L}{\de\phi} 
$$
or, equivalently,
$$\pt_\mu E^\mu\approx 0$$
and calls it a {\it weak conservation law}. It follows
then from the Poincar\'{e} lemma that (at least locally) there exists on shell
a skew-symmetric quantity  $U^{\mu\nu}=-U^{\nu\mu}$, called a
{\it superpotential}, such that
$$E^\mu\approx\pt_\nu U^{\mu\nu}$$
i.e. $E^\mu$ differs from the divergence $\pt_\nu U^{\mu\nu}$ by a quantity
which vanishes on shell. Notice that the superpotential so defined depends
explicitly on field equations. However, if a formula on
$U^{\mu\nu}$ can be "analytically" prolonged to all field
configurations\footnote{This is always the case for so-called local symmetries
(see below).}  then a quantity $e^\mu = \pt_\nu
U^{\mu\nu}$ defined for all field configurations is automatically conserved
also "off shell", i.e. $$\pt_\mu e^\mu \equiv 0$$ independently of the dynamics.
In this case we call this a {\it strong conservation law}.

\section{Second Theorem of Noether and Superpotentials}

In the present paper we shall be interested in so-called {\it local symmetries}
\  (and second	Noether's theorem). This corresponds to the situation when the
variation  $\de_*\phi$ depends on some (finite) number of arbitrary functions
$\th=\{\th^i (x)\}$ $i=1,\ldots,N$ over spacetime $M$ together with their
derivatives up	to some (finite) order, say $s$. In other words one has
$$
\de_*\phi\equiv \de_{\th}\phi = \vp_{i}\th^{i} +
\vp^{\rho}_{i}\pt_{\rho}\th^{i} +\ldots +
\vp^{\rho_1\ldots\rho_s}_{i}\pt_{\rho_1}\ldots\pt_{\rho_s}\th^{i}
\eqno (5)
$$
The order $s$ of the highest derivative involved in this field transformation
law is called the {\it geometric order} of the transformation. If the
Lagrangian  is of order $k$, i.e. it involves up to $k$ derivatives of the
fields, the integer $r=s+k$ is then called the {\it total order} of the theory
\cite{FF1}.

Supposing now
\footnote{We slightly extend here the formalism developed in \cite{FF2},
where proofs of our statements can also be found.}
that the total order of the theory is at most $3$,
then both sides of equation (3) become differential
operators acting on $\th$, where
$$
E^{\mu} \equiv E^{\mu} (\th)= t^{\mu}_{i}\th^{i}+
t^{\mu \rho}_{i}\pt_{\rho}\th^{i} +
t^{\mu \rho \sigma}_{i}\pt_{\rho}\pt_{\sigma}\th^{i}
\eqno (6)
$$
$$
\frac{\delta L}{\delta \phi}\de_\th\phi  \equiv  W (\th) = w_{i}\th^{i}+
w^{\rho}_{i}\pt_{\rho}\th^{i} +
w^{\rho \sigma}_{i}\pt_{\rho}\pt_{\sigma}\th^{i}
\eqno (7)
$$
Notice that in	(6) and (7), in order to preserve the
uniqueness of the decomposition, one should assume the following
symmetry conditions for the highest coefficients
\footnote{We shall use the notation
$A^{[\mu\nu]}=\frac{1}{2}(A^{\mu\nu}-A^{\nu\mu})$ for the skew-symmetrisation
and $A^{(\mu\nu)}=\frac{1}{2}(A^{\mu\nu}+A^{\nu\mu})$ for the symmetrisation.}:
$$
t^{\mu [ \rho \sigma ]}_{i} =0,\
\ w ^{ [ \rho \sigma ]}_{i} = 0  
$$
Replacing now (6) and (7) into equation (3) and calculating the
divergence $\pt_\mu E^\mu$ explicitly, one gets an identity which holds true
for arbitrary functions $\th^i$. Therefore the total coefficients resulting in
front of $\th^i$, $\pt_\al \th^i$, etc..., should vanish independently.
This produces, in fact, a differential indentity involving the
Euler-Lagrange expression, so-called {\it generalized Bianchi identity}
\cite{FF2}, which in this case reads as follows
$$
w_{i} - \pt_{\nu}w^{\nu}_{i} + \pt_{\nu}\pt_{\rho}w^{\nu \rho}
_{i} = 0  \eqno (8)
$$
The coefficients of $W$ as given by (7) vanish on shell  because of the
Euler--\linebreak [2] Lagrange equations, so that $\pt_\mu E^\mu\approx 0$ 
holds. It appears, in this case, that the
Noether current admits a representation of the form
$$
E^{\mu} = \tilde{E}^{\mu} \ +\ \pt_{\rho}U^{\mu \rho}
\equiv \tilde E^\mu  + e^\mu \eqno (9)
$$
where  $\tilde{E}^{\mu}$ is the {\it ``reduced'' Noether current},
which vanishes on shell (see \cite{FF2}). The reduced current
$\tilde E$ is unique, while the superpotential is not, since (9) remains
unchanged if $U^{\mu\nu}$ is redefined by the addition of the divergence of any
skew-symmetric quantity $Y^{\mu\nu\rho}=Y^{[\nu\mu\rho ]}$.  It can be found
that equation (9) has an explicit solution given by
$$
\tilde{E}^{\mu} = (w^{\mu}_{i} - \pt_{\rho}w^{\rho \mu}_{i
}) \th^{i} + w^{\mu \rho}_{i}\pt_{\rho}\th^{i}
\eqno (10)
$$
$$
U^{\mu \rho}=(t^{[ \mu \rho ]}_{i} + \pt_{\nu} \tilde{t}^{\nu [
\mu \rho ]}_{i})\th^{i} + \tilde{t}^{\mu \rho \nu}_{i
}\pt_{\nu}\th^{i}
 \eqno (11)
$$
where $\tilde{t}^{\mu \rho \nu}_{i}= (4/3)t^{[ \mu \rho ] \nu}_{i}$.
This gives explicit formulae for a superpotential in the case of theories
of total order at most $3$. It should be noticed that the "lowest"
coefficients $t^\mu_i$ do not enter this explicit expression (11); therefore,
this suggests that superpotentials are, in a suitable sense, "algebraically
simpler" then the corresponding Noether currents.

We should also remark that formula (11) seems to provide a "canonical" 
expression for the superpotential; but, in general, this is not true. 
In fact, equation (11) explicitly depends on the local coordinate system we 
have chosen as well as the transformation properties of the coefficients 
appearing in (6). Nevertheless, as we shall see below,	such a canonical 
expression will be possible under some additional assumption.

As a general class of examples including theories of gravity (which in
fact we shall need in the sequel) let us assume that the field $\phi$ is
a field of {\it geometric objects} over $M$ \cite{FF1}, i.e. we assume
that  under coordinate changes in $M$ the field  $\phi$
transforms properly  through laws which depend
only on a finite number of the partial derivatives of the coordinate change
itself~\footnote{Tensors or linear connections are geometric objects, while
arbitrary gauge fields are not.}~.
We assume also that the Lagrangian is  {\it natural}  (or, equivalently, {\it
reparametrization invariant}),	in the sense that the lift of all
diffeomorphisms of $M$ transform $L$ as a scalar density of weight $1$.  This
means that, at the infinitesimal level, one has
$$\de_*\phi = {\cal L}_{\xi}\phi \eqno (12)$$ and
$$\de_*L={\cal L}_{\xi}L=\pt_\al(\xi^\al L) \eqno (13)
$$
where $\xi=\xi^\al\pt_\al$ is an arbitrary vectorfield on $M$
(i.e., an infinitesimal diffeomorphism) and ${\cal L}_{\xi}$ denotes the Lie
derivative along $\xi$.

As a consequence, equation  (3) can be written as follows:
$$
\partial_{\mu}E^{\mu}\equiv \pt_\mu [\vr^\mu (\phi, \de_\xi\phi)-\xi^\mu L] =
-\ \frac{\delta L}{\delta \phi} {\cal L}_{\xi}\phi   
$$
where $E^\mu\equiv E^\mu (\xi)$ is the appropriate Noether current evaluated
along the infinitesimal symmetry given by $\xi$, called now
{\it energy--momentum flow}.
It is known that the Lie derivative of any geometric object can be expressed
by means of covariant derivatives with respect to an arbitrary symmetric linear
connection on $M$. In particular, we may expand the Lie
derivative ${\cal L}_{\xi}\phi$ as a linear combination of $\xi$ and
its covariant derivatives, i.e. we write
$$
{\cal L}_{\xi}\phi = \Phi_{\alpha}\xi^{\alpha} +
\Phi^{\rho}_{\alpha}\nabla_{\rho}\xi^{\alpha} +\ldots
\eqno (14) $$
Suppose again
\footnote{This includes all standard metric theories of gravity ($k=2,\ s=1$) 
as well as so-called first-order formalism (or Palatini), with
$k=1,\ s=2$ (Yang-Mills theories have $k=1,\ s=1$).}
that the total order of the theory is at most $3$.
This implies global and covariant decompositions
$$
E^{\mu} \equiv E^{\mu} (\xi)= T^{\mu}_{\alpha}\xi^{\alpha}+
T^{\mu \rho}_{\alpha}\nabla_{\rho}\xi^{\alpha} +
T^{\mu \rho \sigma}_{\alpha}\nabla_{\rho}\nabla_{\sigma}\xi^{\alpha}
\eqno (15) $$
$$
\frac{\delta L}{\delta \phi}\de_\xi\phi  \equiv  W (\xi) =
W_{\alpha}\xi^{\alpha}+ W^{\rho}_{\alpha}\nabla_{\rho}\xi^{\alpha} +
W^{\rho \sigma}_{\alpha}\nabla_{\rho}\nabla_{\sigma}\xi^{\alpha}
\eqno (16) $$
According to the general techniques developed in \cite{FF2}
we get thence covariant versions of equations (9), (10) and (11), i.e.
$$
W_{\mu} - \nabla_{\nu}W^{\nu}_{\mu} + \nabla_{\nu}\nabla_{\rho}W^{\nu \rho}
_{\mu} = 0  \eqno (17)
$$
as a generalized Bianchi identity;
$$
\tilde{E}^{\mu} = (W^{\mu}_{\alpha} - \nabla_{\rho}W^{\rho \mu}_{\alpha
}) \xi^{\alpha} + W^{\mu \rho}_{\alpha}\nabla_{\rho}\xi^{\alpha} \eqno(18)
$$
as a reduced energy flow; and finally
$$
U^{\mu \rho}=(T^{[ \mu \rho ]}_{\alpha} + \nabla_{\nu} \tilde{T}^{\nu [
\mu \rho ]}_{\alpha})\xi^{\alpha} + \tilde{T}^{\mu \rho \nu}_{\alpha
}\nabla_{\nu}\xi^{\alpha}
\eqno (19) $$
where $\tilde{T}^{\mu \rho \nu}_{\alpha}= (4/3)T^{[ \mu \rho ] \nu}_{
\alpha}$ (recall that $T^{\mu [ \rho \nu ]}_{\alpha}=0$).
Equation (19) gives the required formula for  a {\it canonical} superpotential 
in the case of natural theories of total order at most $3$ 
(see \cite{FF2}).

As an example one can consider the superpotential obtained from the
Einstein-Hilbert (purely metric) gravitational Lagrangian
(here notation is standard)
$$
L_{EH}(g, \pt g, \pt^2 g) = |det g|^{\frac{1}{2}} g^{\mu\nu} R_{\mu\nu}
$$
which is given by
$$
U_{EH}^{\mu\nu}(\xi) = |det g|^{\frac{1}{2}}
(\na^\mu\xi^\nu - \na^\nu\xi^\mu)
$$
and it is known as the {\it Komar superpotential}
\footnote{We refere the reader to e.g. \cite{Go,BCJ} for more exhausted 
discussion of the Komar superpotential.}. 
It has been recently shown that the Komar expression is "universal" in the 
following sense: for a large class of non-linear gravitational Lagrangians 
(in the first-order "Palatini" formalism), the superpotential does not depend 
on the Lagrangian and is in fact equal to the  Komar superpotential 
\footnote{ It has been also found that in these cases the universality 
property holds for the Einstein equations as well.}
\cite{BFFV,BFFV2}.

In some mathematical literature  currents admitting superpotentials are
called {\it trivial} since they lead to strong conservation laws, which
hold true irrespectively of the form of the equations of motion \cite{TA}.
Such currents are also somehow "trivial" from a co-cohomological
viewpoint (in the sense the they define trivial co-homology classes
in de Rahm cohomology, being on shell exact forms).
Nevertheless, we remark that they produce physically relevant quantities, 
like e.g. charges, masses and so on. 

The advantage of using superpotentials is twofold. First, as we have
mentioned before, they are in general (algebraically) simpler expressions 
then the corresponding Noether currents. Moreover, in order to calculate the 
flux of any conserved quantity they allow, via Stokes' theorem, to reduce 
calculations to the following:
$$
\int_\Sigma E^\mu d\Sigma_\mu = \oint_{\pt\Sigma}U^{\mu\nu}d\Sigma_{\mu\nu}
$$
Here $\Sigma$ is an hypersurface in $M$ of codimension $1$ and $\pt\Sigma$
denotes its boundary (which has codimension $2$). In the case of $d=4$
spacetime this simply means that the flux of a conserved quantity through
a $3$-dimensional portion of spacetime can be calculated as the integral of
the superpotential along the $2$-dimensional boundary $\pt\Sigma$, which
is a surface.

\section{Chern--Simons Gauge Theory}

To start with, we discuss another physically interesting example, 
i.e. the Chern-Simons theory  considered as a gauge theory over 
$M=\Re^3$ (see also \cite{DJT,BCJ,heh}). The Chern-Simons three-form 
written in terms of a connection and its curvature reads as follows
$$
L^{(3)}_{CS}= \tr (\Om\w\om - \frac{1}{3}\om\w\om\w\om)
\eqno (20)$$
where $\om=\A_\mu dx^\mu$ is a connection one-form, $\A_\mu$ is a
matrix-valued \footnote{In fact, $\A_\mu$ belongs to the Lie algebra of some
gauge  group $G$, which for simplicity we assume here to be represented  in
terms of (complex) matrices.}  gauge potential,
$\Om=d\om+\om\w\om=\frac{1}{2}\F_{\mu\nu}dx^\mu\w dx^\nu$ is  its curvature
two-form, $\F_{\mu\nu}= \pt_\mu \A_\nu-\pt_\nu \A_\mu +  [\A_\mu, \A_\nu]$
is the gauge field strength and $\tr$ denotes the matrix trace.

The corresponding Lagrangian density
$L^{(3)}_{CS}=L_{CS}dx^1\w dx^2\w dx^3$ written in terms of more physically
relevant  quantities is
$$
L_{CS}= \frac{1}{2}\ve^{\mu\nu\rho}\tr (\F_{\mu\nu}\A_\rho - \frac{2}{3}\A_\mu
\A_\nu \A_\rho) \eqno (21)$$
where $\ve^{\mu\nu\rho}$ is the totally skew-symmetric Levi-Civita symbol
in $d=3$.

In this case the first variation formula (1) reads as
$$
\de L_{CS}= \ve^{\mu\nu\rho}\tr (\F_{\mu\nu}\de \A_\rho) - \pt_\mu
[\ve^{\mu\nu\rho} \tr(\A_{\nu}\de \A_\rho)] \eqno (22)$$
and this  gives $\F_{\mu\nu}=0$ as the equation of motion, which means $\Om=0$,
i.e. that only flat connections  are solutions of the equations of motion.

As a symmetry one can consider any infinitesimal gauge transformation
$$\de_\chi \A_\mu=D_\mu\chi \eqno (23)$$
where $\chi$ is an arbitrary (matrix-valued) function and
$D_\mu= \pt_\mu +[\A_\mu,\cdot \,]$ denotes the covariant derivative with
respect  to the connection $\om$.  It is known from transformation properties
of the Lagrangian $L_{CS}$ (see e.g. \cite{DJT,CF}) that the following holds:
$$\de_\chi L_{CS}=\pt_\mu [\ve^{\mu\nu\rho} \tr (\A_{\nu}
\pt_\rho\chi)] \eqno (24)$$
Replacing now (22), (23) and (24) into equation (3) we get
$$
\ve^{\mu\nu\rho}\tr (\F_{\mu\nu}D_\rho\chi) = \pt_\mu E^\mu_{CS}(\chi)
\eqno (25)
$$
where
$$
E^\mu_{CS}(\chi)=2\ve^{\mu\nu\rho}\tr (\A_\nu\A_\rho\chi) +
2\ve^{\mu\nu\rho}\tr (\A_\nu\pt_\rho\chi)=
\eqno (26)$$
$$
-2\ve^{\mu\nu\rho}\tr (\A_\nu\A_\rho\chi) +
2\ve^{\mu\nu\rho}\tr (\A_\nu D_\rho\chi)
$$
is the expression for the Noether current stemming from the gauge 
invariance of  $I_{CS}$.
Therefore, indentifying the corresponding coefficients in (26) with
those in (7), we obtain with the help of (11) the following superpotential
(see also \cite{BCJ}):
$$
U_{CS}^{\mu\rho}(\chi)=2\ve^{\mu\nu\rho}\tr (\A_\nu\chi)
\eqno (27)$$
The left hand side of (25), after expanding it as a differential operator
acting on $\chi$, gives rise (see also equation (8)) to the generalized Bianchi
identity
$$
\ve^{\mu\nu\rho}D_\rho\F_{\mu\nu} = 0 \eqno (28)
$$
which, in dimension $d=3$, is known to be equivalent to the standard
Bianchi identity for the gauge field $\F_{\mu\nu}$, i.e.
$$D_\rho\F_{\mu\nu}+D_\mu\F_{\nu\rho}+D_\nu\F_{\rho\mu}=0$$

A similar analysis can be performed for
the diffeomorphism invariance of $L_{CS}$; in this case one has
\footnote{Under diffeomorphisms the gauge potential $\A_\mu$ behaves 
as a one-form and $L_{CS}$ as a scalar density of weight one.}:
$$\de_\xi\A_\mu\equiv {\cal L}_\xi \A_\mu = \xi^\al\pt_\al\A_\mu + 
\A_\al\pt_\mu\xi^\al \eqno (29)$$ and
$$\de_\xi L_{CS}\equiv {\cal L}_\xi L_{CS} = \pt_\al (\xi^\al L_{CS}) 
\eqno (30)$$
Now we have:
$$
E^\mu(\xi)=\xi^\mu L_{CS} +\ve^{\mu\nu\rho}\tr(\A_\nu\pt_\al\A_\rho)\xi^\al +
\ve^{\mu\nu\rho}\tr(\A_\nu\A_\al)\pt_\rho\xi^\al
\eqno (31)$$
from which it follows via (11)
$$
U_{CS}^{\mu\rho}(\xi)=\ve^{\mu\nu\rho}\tr(\A_\nu\A_\al)\xi^\al
\eqno (32)$$
since only the last term in (31) contributes into (32).

The generalized Bianchi identity takes thence the form
$$
\ve^{\mu\nu\rho}\tr (D_\rho\F_{\mu\nu}\A_\al -
\F_{\mu\nu}\F_{\al\rho}) = 0 \eqno (33)
$$
in which the first term vanishes owing to (28), so that (33) reduces 
in fact to the funny algebraic identity
$$
\ve^{\mu\nu\rho}\tr (\F_{\mu\nu}\F_{\al\rho}) = 0 \eqno (34)
$$
which holds true for {\underline{any}} gauge field in dimension three!

For a comparison, we recall the reader that the gauge and diffeomorphism
invariance of Yang-Mills Lagrangians in four dimensions lead instead to
$$
U_{YM}^{\mu\rho}(\chi)= \tr (\F^{\mu\rho}\chi) \eqno (35)
$$
$$
U_{YM}^{\mu\rho}(\xi)=\tr (\F^{\mu\rho}\A_\al)\xi^\al \eqno (36)
$$

Some autors \cite{Jac,BH} prefer to deal with so called {\it improved
diffeomorphisms}
$$
\hat\de_\xi\A_\mu\equiv \xi^\al\F_{\al\mu}=\de_\xi\A_\mu
- D_\mu(\xi^\al\A_\al) \eqno (37)
$$
which differ from the Lie derivative (29) only by a gauge transformation
$\hat\chi = \xi^\al\A_\al$. Direct calculations show that
$$
\hat U_{CS}^{\mu\rho}(\xi) = U_{CS}^{\mu\rho}(\xi) -
U_{CS}^{\mu\rho}(\hat\chi) 
=\ve^{\mu\rho\nu}\tr(\A_\nu\A_\al)\xi^\al \eqno (38)
$$
i.e., it differs from (32) by a sign. Similarly, one finds from (35-37)
$$
\hat U_{YM}^{\mu\rho}(\xi)=0 \eqno (39)
$$
It means that there are no, in the Yang-Mills theory, non-trivial 
charges corresponding to these transformations.

\section{Chern--Simons Gravity}

We now proceed to discuss a less studied case of Chern-Simons gravity
considered as a purely metric theory over $\Re^3$. Therefore, the 
only dynamical variable is a Riemannian (or pseudo-Riemannian) metric 
$\bg=g_{\mu\nu}$. The Lagrangian density is
$$
L_{CSG} = \frac{1}{2}\ve^{\mu\nu\rho}\tr (\R_{\mu\nu}\G_\rho  -
\frac{2}{3}\G_\mu\G_\nu \G_\rho) $$
$$=\frac{1}{2}\ve^{\mu\nu\rho}(R^\al_{\bt\mu\nu}\Ga^\bt_{\al\rho} -
\frac{2}{3}\Ga^\al_{\bt\mu}\Ga^\bt_{\si\nu}\Ga^\si_{\al\rho})
\eqno (40)$$
where, using the notation introduced in the previous
section, we have now $\om=\G_\mu dx^\mu$, $\G_\mu$ is the matrix-valued
\footnote{Here $\G_\mu\equiv\Ga^\al_{\bt\mu}$ and
$\R_{\mu\nu}\equiv R^\al_{\bt\mu\nu}$ are represented as $3\times 3$
matrices and the upper index is the row index.}
one-form of Riemannian connection,
$\Om=\frac{1}{2}\R_{\mu\nu}dx^\mu\w dx^\nu$ is the Riemann
curvature two-form and $\R_{\mu\nu}= \pt_\mu \G_\nu-\pt_\nu \G_\mu +
[\G_\mu,\G_\nu]$ is the Riemann curvature tensor written in matrix notation.
For the sake of completeness we also recall two basic formulae of Riemannian
geometry
$$
\Ga^\al_{\bt\mu}= \frac{1}{2}g^{\al\si}(\pt_\bt g_{\mu\si} +\pt_\mu
g_{\si\bt} - \pt_\si g_{\bt\mu}) \eqno (41)$$
$$
R^\al_{\bt\mu\nu}=\pt_\mu\Ga^\al_{\bt\nu}-\pt_\nu\Ga^\al_{\bt\mu} +
\Ga^\al_{\si\mu}\Ga^\si_{\bt\nu}-\Ga^\al_{\si\nu}\Ga^\si_{\bt\mu}
\eqno (42)$$
as well as the following indentity
$$
R^\al_{\bt\mu\nu}= \de^\al_\mu R_{\bt\nu} - \de^\al_\nu R_{\bt\mu} +
g_{\bt\nu}R^\al_{\,\mu} - g_{\bt\mu}R^\al_{\,\nu} +
\frac{1}{2}R(g_{\bt\mu}\de^\al_{\,\nu}- g_{\bt\nu}\de^\al_{\,\mu})
\eqno (43)$$
which holds true only in dimension $d=3$ (see e.g. \cite{Yo})
\footnote{This means, in fact, that
in three dimensions the two notions of Ricci flatness and flatness coincide:
i.e., that  gravity is trivial in dimension $d=3$. This is why a coupling
with Chern-Simons term enriches the theory.}.
Here, $R_{\mu\nu}\equiv R^\al_{\mu\al\nu}$ denotes the Ricci tensor and
$R=R^\al_{\,\al}$ is the Ricci scalar
\footnote{We shall be frequently using $\bg$ for raising and lowering
indices.} of $\bg$.

Notice that $L_{CSG}$ depends on the metric $\bg$ only through the
Riemannian connection (41). The relation (22) remains still valid
$$
\de L_{CSG}= \ve^{\mu\nu\rho}\tr (\R_{\mu\nu}\de \G_\rho) - \pt_\mu
[\ve^{\mu\nu\rho} \tr(\G_{\nu}\de \G_\rho)] 
$$
but it does not correspond now to the first variation formula (1), 
since $\G_\rho$ is no longer a dynamical variable in our theory.
Accordingly,  one has to replace $\de\G_\rho$ in the first term 
by means of the well known "Palatini formula"
$$
\de\Ga^\al_{\bt\rho}= \frac{1}{2}g^{\al\si}(\na_\bt\de g_{\rho\si} +
\na_\rho\de g_{\si\bt} - \na_\si\de g_{\bt\rho}) \eqno (44)$$
Making use of equation (43) together with  the contracted Bianchi 
identity $\na_\al R^\al_\mu=\frac{1}{2}\na_\mu R$ one finds
$$
\ve^{\mu\nu\rho}R^\bt_{\al\mu\nu}\de \Ga^\al_{\bt\rho} =
- 2 \ve^{\mu\nu\rho}\na_\mu R^\al_\nu \de g_{\al\rho}
+ \pt_\mu (2\ve^{\mu\nu\rho}R^\al_\nu\de g_{\al\rho})
\eqno (45)$$
With the help of the tricky identity which holds true for any 
quantity $S^\si_{\mu\nu}$ 
$$
2\ve^{\mu\nu [\rho}S^{\si ]}_{\mu\nu} \equiv 
\ve^{\rho\si\bt}(S^\al_{\al\bt}-S^\al_{\bt\al})
\eqno (46)$$
the symmetric part of the first term of the right hand side in (45) 
can be converted into the following form
$$
-2 \ve^{\mu\nu\rho}\na_\mu(R^\al_\nu -
\frac{1}{4}\de^\al_\nu R)\de g_{\al\rho}
$$
Therefore, the first variation formula reads now as
$$
\de L_{CSG}=-2 C^{\al\rho}\de g_{\al\rho} + \pt_\mu[\ve^{\mu\nu\rho}
(2R^\al_\nu\de g_{\al\rho} - \tr (\G_\nu\de \G_\rho))]
\eqno (47)$$
where
$$
 C^{\al\rho}=\ve^{\mu\nu\rho}\na_\mu(R^\al_\nu - \frac{1}{4}\de^\al_\nu R)
\eqno (48)$$
is the so-called {\it Cotton tensor} \footnote{
In fact, $C^{\al\rho}$ is a tensor denity of weight $1$.}
, the vanishing of which gives, in the
case of $L_{CSG}$, the	Euler-Lagrange equations of motion. It is also 
known (see e.g. \cite{DJT,Yo}) that the Cotton tensor is symmetric, 
traceless ($C^\al_\al=0$), divergence-free ($\na_\al C^\al_\bt=0$) and 
it vanishes if and only if the metric $\bg$ is conformally flat 
($\equiv$ conformal to a flat one).

As a symmetry transformation, consider then a $1$-parameter group of
diffeomorphisms generated by the vectorfield $\xi=\xi^\al\pt_\al$. In 
this case one  can use the well known expressions
$$
\de_\xi\bg\equiv {\cal L}_\xi g_{\al\rho}=\na_{\al}\xi_{\rho} +
\na_{\rho}\xi_{\al}
\eqno (49)$$
and
$$
\de_\xi\G_\rho\equiv {\cal L}_\xi \Ga^\bt_{\al\rho}=
\xi^\si R^\bt_{\al\si\rho} +\na_{\al}\na_{\rho}\xi^\bt
\eqno (50)$$
In a Riemannian case we are dealing with, the last formula is, of course,
equivalent to the Palatini one, if one replaces $\de g$ in (44) by (49).

Thus, our task is now to calculate $\de_\xi L_{CSG}$ directly from its
transformation properties. Notice that the transformation rule for the 
linear connection
$$
\bar \Ga^\la_{\nu\mu}=\frac{\pt\bx^\la}{\pt x^\rho}\Ga^\rho_{\al\bt}
\frac{\pt x^\al}{\pt\bx^\nu}\frac{\pt x^\bt}{\pt\bx^\mu}  +
\frac{\pt\bx^\la}{\pt x^\rho}\frac{\pt^2 x^\rho}{\pt\bx^\nu\pt\bx^\mu}
$$
$$
=(\frac{\pt\bx^\la}{\pt x^\rho}\Ga^\rho_{\al\bt}
\frac{\pt x^\al}{\pt\bx^\nu} +\frac{\pt\bx^\la}{\pt x^\rho}
\frac{\pt}{\pt x^\bt}
[\frac{\pt x^\rho}{\pt\bx^\nu}])\frac{\pt x^\bt}{\pt\bx^\mu}
$$
is the composition of a vector- and of a pure gauge transformation.
In other words one has
$$
\bar{\G}_\mu=(\J\G_\bt\J^{-1}+\J\pt_\bt\J^{-1})
\frac{\pt x^\bt}{\pt\bx^\mu}   \eqno (51)$$
where $\J=\frac{\pt\bx^\la}{\pt x^\rho}$ is the Jacobian matrix. Therefore,
at the infinitesimal level ($\bar x^\al=x^\al-\xi^\al$), $\de_\xi L_{CSG}$ 
is the sum of two terms 
\footnote{In fact, we have to substitute $\chi\mapsto\frac{\pt\xi^\bt}
{\pt x^\al}$ in equation (24).}
of the type (24) and (13). More precisely we have obtained
$$
\de_\xi L_{CSG}=\pt_\mu[\xi^\mu L_{CSG} + \ve^{\mu\nu\rho}
\tr (\G_{\nu}\pt_\rho\bchi)]
\eqno (52)$$
where $\bchi=\frac{\pt\xi^\bt}{\pt x^\al}$ is an infinitesimal version
of the inverse Jacobian matrix. We see that the Lagrangian $L_{CSG}$ is 
not natural, but it is in some sense {\it quasi-natural}, i.e. it 
differs from a natural one by a total derivative.

In a similar way one deduces from (51) that the following holds
$$
\de_\xi\G_\rho =\na_\rho\bchi + \xi^\al\pt_\al\G_\rho +
\G_\al\pt_\rho\xi^\al
\eqno (53)$$
as the sum of (23) and (29).
Here $\na_\rho\bchi$ denotes the "formal" covariant differential of 
$\bchi$, formally treated  as a tensor of rank $(1, 1)$, i.e. we set 
$\na_\rho\bchi=\pt_\rho\bchi + [\G_\rho, \bchi]$. Of course, it can 
be directly checked that equations (50) and (53)  are equivalent.

Inserting now all necessary expressions in equation (3) we obtain 
the variation formula under the form
$$
2C^{\al\rho}{\cal L}_\xi g_{\al\rho}=-\pt_\mu E^\mu_{CSG}(\xi) 
\eqno (54) $$
where
$$
 E^\mu_{CSG}(\xi)=\xi^\mu L_{CSG} +\ve^{\mu\nu\rho}\{\tr(\G_\nu\pt_\al
\G_\rho)\xi^\al -2R^\tau_\nu(\de^\si_\tau g_{\rho\al}+\de^\si_\rho
g_{\tau\al})\na_\si\xi^\al +
$$
$$
+\tr(\G_\nu [\G_\rho, \bchi]) + \tr(\G_\nu\G_\al)\pt_\rho\xi^\al +
2\tr(\G_\nu\pt_\rho\bchi)\} \eqno (55)
$$
is the energy-momentum complex for the Chern-Simons gravitational 
Lagrangian (40). 
We are now in position 
to obtain the corresponding superpotential
$U^{\mu\rho}_{CSG}$. Since superpotentials are additive quantities one 
can calculate the contributions from  all terms  of (55) separately. 
Moreover, the first two terms are of the lowest order and  do not 
contribute to the superpotential. 

To obtain the contribution from the third term of (55) one will use the
covariant  formula (19).  First  we rewrite  this term under
the form $T^{\mu\si}_\al\na_\si\xi^\al$,
where $T^{\mu\si}_\al=-2\ve^{\mu\nu\rho}R^\bt_\nu(\de^\si_\bt g_{\rho\al}+
\de^\si_\rho g_{\bt\al})$. Then, by making use of the identity (46),
$T^{[\mu\si ]}_\al$ can be converted into the explicit skew-symmetric form
$\ve^{\mu\si\nu}(3R_{\nu\al}-Rg_{\nu\al})$. This gives rise to:
$$
^3U^{\mu\rho}_{CSG}=\ve^{\mu\nu\rho}(Rg_{\nu\al}-3R_{\nu\al})\xi^\al
\eqno (56)$$

Similarly, contributions coming from the fourth and the fifth terms are
easy to calculate directly by using (11) and (46):
$$
^4U^{\mu\rho}_{CSG}+\, ^5U^{\mu\rho}_{CSG}=\ve^{\mu\nu\rho}\{
[\G_\si, \G_\nu]^\si_\al + \tr(\G_\nu\G_\al)\}\xi^\al
\eqno (57)$$
where $[\G_\si, \G_\nu]^\si_\al=\Ga^\si_{\bt\si}\Ga^\bt_{\al\nu}
-\Ga^\si_{\bt\nu}\Ga^\bt_{\al\si}$, which due to the symmetry of
$\Ga^\bt_{\al\si}$ reduces (57) to the form:
$$
^4U^{\mu\rho}_{CSG}+\, ^5U^{\mu\rho}_{CSG}=\ve^{\mu\nu\rho}
(\Ga^\si_{\bt\si}\Ga^\bt_{\al\nu})\xi^\al
\eqno (58)$$

The sixth and last term in (55) is of the second order. We write it as
$t_\al^{\mu\rho\si}\pt_\rho\pt_\si\xi^\al$, with
$t_\al^{\mu\rho\si}=2\ve^{\mu\nu (\rho}\Ga^{\si )}_{\al\nu}$. One needs 
the shorthand
$\tilde t_\al^{\mu\rho\si}=4/3t_\al^{[\mu\rho ]\si}$ (see (11)); 
a direct calculation with the help of another tricky identity
$$
2\ve^{\mu\nu [\rho}A^{\si ]}_\nu \equiv \ve^{\rho\si\nu}(A^\mu_\nu
-\de^\mu_\nu A^\bt_\bt)
\eqno (59)$$
gives then rise to
$\tilde t_\al^{\mu\rho\si}=2\ve^{\mu\nu\rho}\Ga^\si_{\al\nu}+
(2/3)\ve^{\mu\rho\si}\Ga^\bt_{\al\bt}$. Since 
$^6U^{\mu\rho}_{CSG}=\pt_\nu\tilde t_\al^{\nu [\mu\rho]}\xi^\al +
\tilde t_\al^{\mu\rho\si}\pt_\si\xi^\al$, one gets:
$$
^6U^{\mu\rho}_{CSG}=\frac{1}{3}\ve^{\mu\nu\rho}(\pt_\nu\Ga^\bt_{\al\bt}-
3\pt_\bt\Ga^\bt_{\al\nu})\xi^\al + \frac{2}{3}\ve^{\mu\nu\rho}
(3\Ga^\si_{\al\nu}-\de^\si_\nu\Ga^\bt_{\al\bt})\pt_\si\xi^\al
\eqno (60)$$

We can now formulate our main result:
the superpotential $U^{\mu\rho}_{CSG}(\xi)$ appears as the sum:
$$
U^{\mu\rho}_{CSG}(\xi)=\,
^3U^{\mu\rho}_{CSG}+\, ^4U^{\mu\rho}_{CSG}+\, ^5U^{\mu\rho}_{CSG}+\, 
^6U^{\mu\rho}_{CSG}
\eqno (61)$$
with addenda given by (56), (58) and (60).

The left hand side of (54) taking into account (49) has the form
$4C^\al_\bt\na_\al\xi^\bt$. This, of course, leads via (17) to
a divergence-free property of the Cotton tensor as a generalized
Bianchi identity.

\section*{Acknowledgments}

One of us (A.B.) gratefully acknowledges the hospitality of
the Department of Mathematics
of the University of Torino under the support of G.N.F.M. of
Italian C.N.R. This work is sponsored by G.N.F.M., M.U.R.S.T. (40\% \ Proj.
``Metodi Geometrici e Probabilistici in Fisica Matematica''); one of us
(A. B.) acknowledges also support from KBN 2 P302 023 07.


\begin{thebibliography}{}

\bibitem{CS}
S. S. Chern and J. Simons,
{\it Proc. Nat. Acad. Sci. USA} {\bf 68}(4) (1971), 791 \\
$~$\ \ \ S. S. Chern and J. Simons,
{\it Ann. Math.} {\bf 99} (1974), 48

\bibitem{DJT}
S. Deser, R. Jackiw and S. Templeton,
{\it Phys. Rev. Lett.} {\bf 48} (1982), 975\\
$~$\ \ \ S. Deser, R. Jackiw and S. Templeton, {\it Ann. of Physics\/}
{\bf 140} (1982), 372

\bibitem{des}
S. Deser,
{\it Chern-Simons Terms as an Example of the Relations Between 
Mathematics and Physics}, (math-ph/9805020)

\bibitem{Wi}
E. Witten,
{\it Commun. Math. Phys.}  {\bf 121} (1989),  351

\bibitem{An}
I. M. Anderson, {\it Ann. of Math.\/}, {\bf 120} (1984) 329

\bibitem{CF}
A. H. Chamseddine and J. Fr\"ohlich,
{\it Commun. Math. Phys.} {\bf 147} (1992), 549

\bibitem{Go}
J. N. Goldberg, {\it Phys. Rev.} {\bf 111}\, (1958),  315\\
$~$\ \ \ J. N. Goldberg, {\it Phys. Rev. D}{\bf 41}(2)\, (1990),  410

\bibitem{Fl}
J. G. Fletcher, {\it Rev. Mod. Phys.} {\bf 32} (1) (1960),  65

\bibitem{BCJ}
D. Bak, D. Cangemi and R. Jackiw,
{\it Phys. Rev. D}{\bf 49} (1994), 5173

\bibitem{heh} 
F. W. Hehl, J. D. McCrea, E. W. Mielke and Y. Ne'eman, 
Phys. Rep. {\bf 258} (1995), 1

\bibitem{JS} 
B. Julia and S. Silva, 
Class. Quantum Grav. {\bf 15} (1998), 2173  \\
S. Silva, {\it On Superpotentials and Charge Algebras of
Gauge Theories} -- hep-th/9809109

\bibitem{FF1}
M. Ferraris and M. Francaviglia,
{\it Journ. Math. Phys.} {\bf 26}(6)\, (1985), 1243

\bibitem{FF2}
M. Ferraris, M. Francaviglia and O. Robutti,
in:   {\it G\'eom\'etrie et
Physique (Proceedings  Journe\`ees Relativistes  de Marseille 1985)};
Y. Choquet-Bruhat, B. Coll, R. Kerner and A. Lichnerowicz  eds.;
Travaux en Cours, Hermann (Paris, 1987),  pp. 112-125\\
M. Ferraris and M. Francaviglia,
{\it Class. Quantum Grav.}  {\bf 9}, (Supplement 1992),  S79

\bibitem{FFF}
L. Fatibene, M. Ferraris and M. Francaviglia,
{\it Journ. Math. Phys.} {\bf 38}(8)\, (1997),  3953

\bibitem{sar}
G. Sardanashvily, Class. Quantum Grav. {\bf 14}\, (1997), 1371 \\
G. Giachetta, L. Mangiarotti and G. Sardanashvily, 
{\it Energy-Momentum and Gauge Conservation Laws}, (gr-qc/9807054)

\bibitem{BFFV}
A. Borowiec, M. Ferraris, M. Francaviglia and I. Volovich,
{\it Gen. Rel. Grav.} {\bf 26}(7) (1994), 637

\bibitem{BFFV2}
A. Borowiec, M. Ferraris, M. Francaviglia and I. Volovich,
{\it Class. Quantum Grav.} {\bf 15}(1) (1998), 43 (gr-qc/9611067)\\
A. Borowiec and M. Francaviglia,
{\it Alternative Lagrangians for Einstein Metrics}, Proc. Int.
Sem. Math. Cosmol., Potsdam 1998, M. Rainer and H.-J. Schmidt
(eds.), WSPC - to appear; (gr-qc/9806116)

\bibitem{TA}
C. G. Torre and I. M. Anderson,
{\it Phys. Rev. Lett.} {\bf 70}(23)\, (1993), 3525

\bibitem{Yo}
J. W. York,
{\it Phys. Rev. Lett.} {\bf 26}(26)\, (1971), 1656

\bibitem{Jac}
R. Jackiw,
{\it Phys. Rev. Lett.} {\bf 41}(24)\, (1978), 1635

\bibitem{BH}
M. Ba$\tilde{n}\!$ados L. J. Garay and M. Henneaux,
{\it Phys. Rev. D}{\bf 53}(2) (1996), R593\\
M. Ba$\tilde{n}\!$ados, M. Henneaux, C. Iannuzzo and C. M. Viallet,
{\it Class. Quantum Grav.} {\bf 14} (1997), 2455 

\end{thebibliography}
\end{document}